\documentclass[twocolumn,showpacs,nofootinbib,preprintnumbers,amsmath,amssymb]{revtex4}

\usepackage[dvips]{graphicx}

\begin{document}

\title{Automatic Control over the Cosmological Constant \\ through Non-minimal Phantom and Quintessence}

\author{Je-An~Gu}
\email{jagu@phys.cts.nthu.edu.tw} %
\affiliation{National Center for Theoretical Sciences, Hsinchu 30013, Taiwan, R.O.C.} %

\begin{abstract}
A mechanism to control the cosmological constant through a scalar
field non-minimally coupled to gravity is proposed. By utilizing
non-minimal phantom or quintessence, the cosmological constant,
which may be large originally, can be automatically driven to a
value on the scale of the mass parameter in the
phantom/quintessence potential $V(\phi)$. The reduction of a large
cosmological constant involves the weakening of gravity that
therefore may be much stronger initially. There exist the cases
where originally gravity is on the TeV scale so that the hierarchy
between gravity and three gauge interactions in the standard model
of particle physics is bridged at the beginning. Although the
cosmological constant can be automatically tuned or largely
reduced under this mechanism, its energy density may still remain
on the same order of magnitude as the original one. Thus,
explaining the smallness of the observation-suggested cosmological
constant energy density is still a difficult mission yet to be
completed.
\end{abstract}


\maketitle


\textit{Introduction.} The existence of a small positive
cosmological constant (CC) or, more conservatively, an upper limit
to the CC is indicated by a variety of astrophysical observations
\cite{Perlmutter:1998np,Riess:1998cb,SNLS,Riess:2006fw,ESSENCE:2007jb,WMAP:2006,Tegmark:2006}.
This observation-suggested CC (upper limit) is on the scale $H_0
\sim 10^{-33}\textrm{eV}$, i.e.\ with the energy density on the
scale $\sim 10^{-3}\textrm{eV}$. In contrast, a large CC from
vacuum energy is expected in the framework of quantum field
theory, whose energy density is on the scale of the quantum
fluctuations under consideration. If this vacuum energy exists and
does gravitate, even the quantum fluctuations on the micron scale
can have ruined our universe in the early time, needless to say
the Planck scale $M_\textrm{Pl}$ or the supersymmetry breaking
scale $M_\textrm{SUSY}$ that may be the relevant scale for the
vacuum energy. This contrast between the very large and the very
small CC is a long-standing issue called ``cosmological constant
problem'' \cite{Weinberg:1988cp}. For a review of possible
approaches to solving this problem, see
Refs.~\onlinecite{Padmanabhan:2002ji} and
\onlinecite{Nobbenhuis:2004wn} and references therein.

The large CC from vacuum energy might be cancelled in a
brute-force manner by finely tuning, for example, the size of the
bare CC. As another example of fine tuning, in the Randall-Sundrum
(RS) scenario \cite{RS:1999} of extra dimension, by requiring a
relation among the CC in the bulk and those on the branes, the
effect of these CCs on the evolution of space-time can be
cancelled, thereby allowing the existence of a static metric
solution, even though all the CCs may be on a very large scale
such as the Planck scale $M_\textrm{Pl}$.

Although such cancellation is artificial, unnatural and therefore
not beautiful to the physicists full of the sense of beauty, no
physical law forbids the creator from exploiting this brute-force
cancellation to create a comfortable universe for human beings to
reside in. Nevertheless, even accepting this not so lovely
cancellation, one still has difficulty of having a small CC, as
described in the following.

In addition to the difficulty from the contrast between the very
large and the very small CC mentioned above, the possible change
of the vacuum energy during the phase transition associated with
spontaneous symmetry breaking (SSB) makes this issue severe. Even
if one obtains a small CC before the SSB phase transition with a
delicate device, after the phase transition the CC energy density
may drop for a certain amount on the scale of the phase
transition, e.g.\ $\sim 300\,$GeV for the electroweak symmetry
breaking, thereby ruining the earlier (nearly) perfect
cancellation. To have perfect cancellation after the phase
transition(s), the creator must foresee all possible phase
transitions and know the very details of the amount of the vacuum
energy density change during each of them, as detailed as
$10^{-3}$eV at least. Then, the creator needs to make the CC
cancellation before the phase transition imperfect, with the
energy density deficit on the scale of the phase transition with
the precision $10^{-3}$eV or better. This fine tuning as a mission
impossible for the creator of our universe plays a crucial part of
the CC problem.

To overcome this obstacle, it will be perfect if the CC can be
controlled automatically to a value one needs. That is, no matter
how large the original CC and the CC changes at some later times
were, the CC would eventually come to the required value. In the
present article a mechanism to control the CC in such a way is
explored. This mechanism is played by a scalar field, phantom
(with negative kinetic energy) or quintessence (with ordinary
positive kinetic energy), which is non-minimally coupled to
gravity. As going to be presented, through this mechanism the
original bare CC under control can be arbitrarily large and its
sign can be either positive or negative. The sign of the resulting
CC can be either positive or negative with the use of non-minimal
phantom, while it is always negative with non-minimal quintessence
\cite{Gu:2001rr}. As to the size of the resulting CC, stabilizing
the CC around a moderate or a large value is achievable, which may
be helpful to the construction of the models involving larger CC,
such as the RS model. Nevertheless, controlling the CC to a small
value, e.g.\ with the energy density on the scale of $10^{-3}$eV
as suggested by observations, is still a task yet to be done.

In this automatic CC control mechanism, the reduction of a large
bare CC involves the weakening of gravity. Accordingly, the
gravity before the performance of this mechanism to reduce the CC
should be stronger. This may address the hierarchy problem
regarding the contrast between gravity on the Planck scale,
$M_\textrm{Pl}$, and three fundamental gauge interactions in the
standard model of particle physics characterized by the scale
$M_\textrm{SM} \sim \textrm{TeV}$, which is 16 orders of magnitude
smaller than $M_\textrm{Pl}$. As going to be shown, there do exist
the cases where originally the gravity is on the TeV scale,
thereby bridging this hierarchy at the very beginning.


\textit{Analysis.} Consider a real scalar field $\Phi$ which is
non-minimally coupled to gravity, as described by the following
action with a power-law potential.
\begin{eqnarray}
S &=& - \frac{1}{2\kappa} \int d^4 x \sqrt{g} \left( \mathcal{R} +
2 \Lambda_0 \right) \nonumber \\
&& + \int d^4 x \sqrt{g} \left[ \frac{S_K}{2} g^{\mu \nu} \left(
\partial_{\mu} \Phi \right) \left( \partial_{\nu} \Phi \right) -
V_{\mathcal{R}}(\Phi) \right] , \\
&& \hspace{-1em} V_{\mathcal{R}} (\Phi) = \frac{1}{2} \xi
\mathcal{R} \Phi^2 + V(\Phi) \, , \\
&& \hspace{-1em} V(\Phi) = S_V M^{4-n} \Phi^n \, , \; M>0 \, . %
\end{eqnarray}
Here $\kappa$ is the gravitational constant characterizing the
original strength of gravity, $\mathcal{R}$ the Ricci scalar,
$\Lambda_0$ the original bare cosmological constant, $\xi$ the
non-minimal coupling constant, and $S_K$ and $S_V$ denote the sign
of the kinetic energy and that of the potential energy (for
positive $\Phi^n$), respectively. Accordingly, $S_K = +1$
corresponds to quintessence and $S_K=-1$ to phantom.

The scalar field equation and the Einstein equations corresponding
to the above action are as follows.
\begin{widetext}
\begin{eqnarray}
0 &=& S_K \Phi^{;\alpha}_{\;\;\, ;\alpha} + \partial_{\Phi}
V_{\mathcal{R}}(\Phi) = S_K \Phi^{;\alpha}_{\;\;\, ;\alpha} +
\left[ \xi \mathcal{R} \Phi + n S_V M^{4-n} \Phi^{n-1} \right] \, , %
\label{eq:field equation of Phi} \\
G_{\mu \nu} &=& \Lambda^{\prime} g_{\mu \nu} + \kappa
_\textrm{eff} \left( T^\textrm{R}_{\mu \nu} + T^\textrm{M}_{\mu \nu}  \right) %
+ \kappa _\textrm{eff} \left[ %
S_K \left( \partial _{\mu}\Phi \right) \left( \partial _{\nu}\Phi
\right) - \mathcal{L}^{\left( 0 \right) }_{\Phi } g_{\mu \nu} %
+ \xi \left( \Phi ^{2} \right)_{;\mu ;\nu} %
- \xi \left( \Phi ^{2} \right)^{;\alpha}_{\;\;\,
;\alpha} g_{\mu \nu} \right] \, , %
\label{eq:modified Einstein equation} \\
&& \mathcal{L}^{\left( 0 \right) }_{\Phi } = %
\frac{1}{2} S_K g^{\alpha \beta} \left( \partial_{\alpha} \Phi
\right) \left( \partial_{\beta} \Phi \right) - V(\Phi) \, , %
\end{eqnarray}
\end{widetext}
where the semicolon `;' denotes the covariant derivative,
$T^\textrm{R}_{\mu \nu}$ and $T^\textrm{M}_{\mu \nu}$ are the
energy contributions from radiation and (pressureless) matter,
respectively, and the ``modified cosmological constant''
$\Lambda^{\prime}$ and the ``effective gravitational constant''
$\kappa _\textrm{eff}$ are defined as follows.
\begin{eqnarray}
\Lambda^{\prime}(\Phi) &\equiv& \frac{\Lambda_0}{1+\kappa \xi \Phi^{2}} \, , \\
\kappa_\textrm{eff}(\Phi) &\equiv& \frac{\kappa}{1+\kappa \xi \Phi^{2}} \, . %
\label{eq:modified Lambda & eff kappa}
\end{eqnarray}
For the present time $\kappa_\textrm{eff} = 8\pi G_N$, where $G_N$
is the Newtonian gravitational constant. The contraction of Eq.\
(\ref{eq:modified Einstein equation}) gives
\begin{eqnarray}
\mathcal{R} &=& -4\Lambda^{\prime} - \kappa_\textrm{eff} \left[
\rho_\textrm{m} + 4 V - S_K \Phi ^{;\alpha} \Phi _{;\alpha} - 3
\xi \left( \Phi^{2} \right) ^{;\alpha}_{\;\;\, ;\alpha} \right] \nonumber %
\label{eq:Ricci scalar} \\
&=& - 4 \Lambda_\textrm{eff} - \kappa_\textrm{eff} \rho_\textrm{m}
\quad \quad \textrm{for constant $\Phi$}  \, , %
\label{eq:Ricci scalar for constant Phi}
\end{eqnarray}
where $\rho_\textrm{m}$ is the 
matter energy density and the ``effective cosmological constant''
$\Lambda_\textrm{eff}$ is defined as
\begin{equation}
\Lambda_\textrm{eff}(\Phi) \equiv \Lambda^{\prime}(\Phi) +
\kappa_\textrm{eff}(\Phi) V(\Phi) = \frac{\Lambda_0 + \kappa S_V
M^{4-n} \Phi^n}{1+\kappa \xi \Phi^2} \, .
\end{equation}
Note that the radiation energy makes no contribution to the Ricci
scalar $\mathcal{R}$.

In the following we investigate the structure of the non-minimal
potential $V_{\mathcal{R}}(\Phi)$ --- finding extrema and
exploring their stability. The equation $\partial_{\Phi}
V_{\mathcal{R}}(\Phi_m) = 0$ gives the location of extrema:
\begin{equation}
- \xi \left[ 4 \Lambda_\textrm{eff}(\Phi_m) +
\kappa_\textrm{eff}(\Phi_m) \rho_\textrm{m} \right] \Phi_m + n S_V
M^{4-n} \Phi_m^{n-1} = 0 \, ,
\end{equation}
or, equivalently,
\begin{eqnarray}
\Phi_m \left[ 4\Lambda_0 + \kappa \rho_\textrm{m} +
(4-n)\kappa S_V M^{4-n} \Phi_m^n \right. \hspace{3em} \nonumber \\
\left. - nS_V M^{4-n} \Phi_m^{n-2}/\xi \right] = 0 \, .
\end{eqnarray}
For $n>1$ there is an extremum at $\Phi_m=0$. In addition,
according to the above equation, there may exist another extremum
at $\Phi_m = v \neq 0$, for which we have
\begin{equation}
\Lambda_\textrm{eff}(v) = \frac{1}{4\xi} n S_V M^{4-n} v^{n-2} -
\frac{1}{4} \kappa_\textrm{eff}(v) \rho_\textrm{m} \, .
\end{equation}

The model with $n=2$ is particularly interesting and will be the
focus in the rest of this article. For $n=2$,
\begin{equation}
\Lambda_\textrm{eff}(v) = \frac{1}{2\xi} S_V M^{2} - \frac{1}{4}
\kappa_\textrm{eff}(v) \rho_\textrm{m} \, ,
\end{equation}
and accordingly the dependence of the resulting effective CC,
$\Lambda_\textrm{eff}(v)$, on the original bare CC, $\Lambda_0$,
is only through $\kappa_\textrm{eff}(v)$ appearing in the second
term in the right-hand side. Furthermore, when the contribution
from the scalar field potential dominates over the matter
contribution, remarkably, the effective CC,
$\Lambda_\textrm{eff}$, is independent of the original bare CC,
$\Lambda_0$, and is dictated simply by $M$ (provided that $\xi$ is
of order unity):
\begin{equation}
\Lambda_\textrm{eff} \cong \frac{1}{2\xi} S_V M^{2} \sim \pm M^2
\quad \textrm{as} \quad \rho_\textrm{m} \ll \left|
\frac{M^2}{\kappa_\textrm{eff}\xi} \right| \, .
\end{equation}
That is, in this case, even if the magnitude of the bare CC,
$\Lambda_0$, is very large originally, the scalar field $\Phi$
will evolve toward $v$ at which the CC is balanced in such a way
that the scale of the resulting effective CC,
$\Lambda_\textrm{eff}$, is characterized by the mass parameter $M$
in the potential, provided that $\Phi_m = v$ stands for a stable
extremum (which will be investigated later). On the other hand,
when the matter contribution dominates, the energy density
associated with the effective CC follows the matter density with
an opposite sign:
\begin{equation}
\rho_{\Lambda_\textrm{eff}} \equiv \Lambda_\textrm{eff} /
\kappa_\textrm{eff} \cong -\frac{1}{4} \rho_\textrm{m} \quad
\textrm{as} \quad \rho_\textrm{m} \gg \left|
\frac{M^2}{\kappa_\textrm{eff}\xi} \right| \, .
\end{equation}

Now we investigate the stability of extrema and explore the
condition for the existence of a stable extremum at $\Phi_m = v
\neq 0$, meanwhile fulfilling $\kappa_\textrm{eff}(v) = 8 \pi G_N
> 0$. The following is the relevant information for the $n=2$ model.
\begin{eqnarray}
\hspace*{-1.5em} \Phi_m = 0 \, & \textrm{or} & \, %
\Phi_m^2 = \frac{1}{\kappa \xi} - \frac{4\Lambda_0 + \kappa
\rho_\textrm{m}}{2 S_V \kappa M^2} \equiv v^2 \, , \\
\hspace*{-1.5em} \kappa_\textrm{eff}(0) = \kappa & \textrm{,} &
\kappa_\textrm{eff}(v) = \frac{2S_V \kappa M^2}{4S_V
M^2 - 4\xi \Lambda_0 - \kappa \xi \rho_\textrm{m}} \, , \\
\hspace*{-1.5em} \Lambda_\textrm{eff}(0) = \Lambda_0 & \textrm{,}
& \Lambda_\textrm{eff}(v) = \frac{1}{2\xi}S_V M^2 - \frac{1}{4}
\kappa_\textrm{eff}(v) \rho_\textrm{m} \, .
\end{eqnarray}
For $\Phi = \Phi_m + \delta \Phi$, the perturbation $\delta \Phi$
satisfies the following field equation.
\begin{equation}
S_K \left( \delta \Phi \right)^{;\alpha}_{\;\;\, ;\alpha} %
- \left( 4 \xi \Lambda_0 + \kappa \xi \rho_\textrm{m} - 2 S_V M^2
\right) \delta \Phi = 0 \textrm{ as }  \Phi_m = 0 ,
\end{equation}
\begin{equation}
S_K^{\prime} %
\left( \delta \Phi \right)^{;\alpha}_{\;\;\, ;\alpha} %
- 4 S_V \kappa_\textrm{eff}(v) \xi M^2 v^2 \delta \Phi = 0 \;
\textrm{ as } \Phi_m = v \, ,
\end{equation}
where
\begin{equation}
S_K^{\prime} \equiv S_K + 6 \kappa_\textrm{eff}(v) \xi^2 v^2 \, ,
\end{equation}
which is always positive for quintessence 
when $\kappa_\textrm{eff}(v)>0$, but can be either positive or
negative for phantom. For the stable extremum at $\Phi_m = v$ to
exist in our universe where $\kappa_\textrm{eff}(v) = 8 \pi G_N$,
it is required that
\begin{equation} \label{eq:stability condition}
\kappa_\textrm{eff}(v) > 0 \, , \; v^2 > 0 \, , \; S_K^{\prime}
S_V \xi < 0 \, .
\end{equation}

In terms of the new variables,
\begin{equation}
A_0 \equiv \Lambda_0 + \frac{1}{4}\kappa \rho_\textrm{m} \, , \; %
A_M \equiv \frac{S_V M^2}{2\xi} \, , \; %
y \equiv \frac{2(S_K + 3\xi)}{S_K + 6\xi} \, ,
\end{equation}
the formulae for the quantities involved in the requirement in
Eq.\ (\ref{eq:stability condition}) can be rewritten as follows.
\begin{eqnarray}
v^2 &=& \frac{1}{\kappa \xi} \left( 1 - \frac{A_0}{A_M} \right) \, , \\
\kappa_\textrm{eff}(v) &=& \frac{\kappa}{2-A_0/A_M} \, , \\
S_K^{\prime} &=& (S_K + 6\xi) \left( \frac{A_0/A_M - y}{A_0/A_M - 2} \right) . %
\end{eqnarray}
When the contribution from the matter density is negligible, $A_0$
represents the original bare CC that would be driven to the
effective one of the size $\simeq A_M$. Accordingly, the ratio
$A_0/A_M$ appearing in the above equations represents the extent
of the CC tuning or reduction that is considered huge in the CC
hierarchy problem.


\textit{Results.} In Table \ref{table:all cases}, we exhaust all
cases through exploring various ranges or values of $S_K$, $S_V$,
$\xi$ and $A_0/|A_M|$, while leaving the sign of $\kappa$
unspecified. The cases where the requirement in Eq.\
(\ref{eq:stability condition}) for the existence of a stable
extremum with positive $\kappa_\textrm{eff}$ is fulfilled are
marked by the symbol ``\textsf{O}'', and the other cases marked by
``\textsf{X}''. These feasible cases marked by ``\textsf{O}'' are
singled out and listed in Table \ref{table:feasible cases}. The
name of each feasible case is given in the first column ``ID''
that contains the information about the sign or the value of
$S_K$, $\kappa$, $S_V$ and $\xi$ which are relevant quantities for
specifying our model. The sign of the resulting effective CC and
the extent of the automatic CC tuning in these feasible cases are
presented in the columns $A_M$ and $A_0/|A_M|$. The feasible cases
marked by the star symbol ``$\star$'' are particularly interesting
because the original bare CC under control in these cases can be
arbitrarily large in its magnitude. Note that in the feasible
quintessence cases the resulting effective CC is always negative,
while in the case of phantom, with different settings, the
resulting effective CC, as well as the original bare CC under
control, can be either positive or negative. (For example,
$P_{++2}$/$P_{+-3}$ can drive an arbitrarily large
negative/positive CC to a positive/negative value.)



\begin{table*}
\caption{\label{table:all cases} All cases (including the feasible
cases and others), presented in two tables: the left table for
quintessence and the right for phantom. The left part of each
table contains the information about the range of $\xi$ and $A_0$
(in unit of $|A_M|$) that, as well as $\{S_K,S_V,\kappa\}$, are
relevant quantities for specifying our model. In the right part
the sign of the quantities involved in the requirement in Eq.\
(\ref{eq:stability condition}) are presented, where $\pm \kappa$
denotes the sign which is the same as or opposite to that of
$\kappa$. When the signs in the columns $v^2$ and
$\kappa_\textrm{eff}(v)$ are the same and that in the column
$-S_K^{\prime} S_V \xi$ is positive, the extent of the automatic
CC tuning indicated in the column $A_0/|A_M|$ can be achieved as
long as the sign of $\kappa$ is specified in the way such that
$v^2$ and $\kappa_\textrm{eff}(v)$ are both positive. These
feasible cases are marked by the symbol ``\textsf{O}'' in the last
column, while the others marked by ``\textsf{X}''. %
[The feasibility of the cases where $\xi = -1/6$ (for
quintessence) or $1/6$ (for phantom) can also be read from this
table, with the help of the following information: for
quintessence, when $\xi=-1/6 \, ^{\pm}$, $y=\pm \infty$; for
phantom, when $\xi=1/6 \, ^{\pm}$, $y=\mp \infty$.]}
\begin{tabular}{cccc|ccccc}
\multicolumn{9}{l}{\hspace{-0.5em} Quintessence ($S_K = +1$)} \vspace{0.3em} \\
\multicolumn{9}{l}{$S_V = +1$} \\
$\xi$ & $A_M$ & $y$ & $A_0/|A_M|$ & $S_K^{\prime}$ &
$v^2$ & $\kappa_\textrm{eff}(v)$ & $-S_K^{\prime} S_V \xi$ &  \\ %
\hline
$<-\frac{1}{6}$ & \ $-$ \ &$<1$&$<-2$& $-$ & $+\kappa$ & $-\kappa$ & $-$ & \textsf{X} \\
        &             & &  $(-2,-1)$ & $+$ & $+\kappa$ & $+\kappa$ & $+$ & \textsf{O} \\
        &             & &  $(-1,-y)$ & $+$ & $-\kappa$ & $+\kappa$ & $+$ & \textsf{X} \\
        &             & &      $>-y$ & $-$ & $-\kappa$ & $+\kappa$ & $-$ & \textsf{X} \\
\hline
$(-\frac{1}{6},0)$ & \ $-$ \ &$>2$&$<-y$&$+$&$+\kappa$ & $-\kappa$ & $+$ & \textsf{X} \\
        &             & &  $(-y,-2)$ & $-$ & $+\kappa$ & $-\kappa$ & $-$ & \textsf{X} \\
        &             & &  $(-2,-1)$ & $+$ & $+\kappa$ & $+\kappa$ & $+$ & \textsf{O} \\
        &             & &      $>-1$ & $+$ & $-\kappa$ & $+\kappa$ & $+$ & \textsf{X} \\
\hline
$>0$    &  \ $+$ \  & $(1,2)$ & $<1$ & $+$ & $+\kappa$ & $+\kappa$ & $-$ & \textsf{X} \\
        &             & &    $(1,y)$ & $+$ & $-\kappa$ & $+\kappa$ & $-$ & \textsf{X} \\
        &             & &    $(y,2)$ & $-$ & $-\kappa$ & $+\kappa$ & $+$ & \textsf{X} \\
        &             & &       $>2$ & $+$ & $-\kappa$ & $-\kappa$ & $-$ & \textsf{X} \\
\multicolumn{9}{l}{$S_V = -1$} \\
$\xi$ & $A_M$ & $y$ & $A_0/|A_M|$ & $S_K^{\prime}$ &
$v^2$ & $\kappa_\textrm{eff}(v)$ & $-S_K^{\prime} S_V \xi$ &  \\ %
\hline
$<-\frac{1}{6}$ & \ $+$ \ &$<1$& $<y$& $-$ & $-\kappa$ & $+\kappa$ & $+$ & \textsf{X} \\
        &             & &    $(y,1)$ & $+$ & $-\kappa$ & $+\kappa$ & $-$ & \textsf{X} \\
        &             & &    $(1,2)$ & $+$ & $+\kappa$ & $+\kappa$ & $-$ & \textsf{X} \\
        &             & &       $>2$ & $-$ & $+\kappa$ & $-\kappa$ & $+$ & \textsf{X} \\
\hline
$(-\frac{1}{6},0)$ & \ $+$ \ &$>2$&$<1$&$+$& $-\kappa$ & $+\kappa$ & $-$ & \textsf{X} \\
        &             & &    $(1,2)$ & $+$ & $+\kappa$ & $+\kappa$ & $-$ & \textsf{X} \\
        &             & &    $(2,y)$ & $-$ & $+\kappa$ & $-\kappa$ & $+$ & \textsf{X} \\
        &             & &       $>y$ & $+$ & $+\kappa$ & $-\kappa$ & $-$ & \textsf{X} \\
\hline
$>0$    &  \ $-$ \ & $(1,2)$ & $<-2$ & $+$ & $-\kappa$ & $-\kappa$ & $+$ & \textsf{O} \\
        &             & &  $(-2,-y)$ & $-$ & $-\kappa$ & $+\kappa$ & $-$ & \textsf{X} \\
        &             & &  $(-y,-1)$ & $+$ & $-\kappa$ & $+\kappa$ & $+$ & \textsf{X} \\
        &             & &      $>-1$ & $+$ & $+\kappa$ & $+\kappa$ & $+$ & \textsf{O} \\ %
\vspace{-1.8em}
\end{tabular}
\hspace{1em}
\begin{tabular}{cccc|ccccc}
\multicolumn{9}{l}{\hspace{-0.5em} Phantom ($S_K = -1$)} \\
\multicolumn{9}{l}{$S_V = +1$} \\
$\xi$ & $A_M$ & $y$ & $A_0/|A_M|$ & $S_K^{\prime}$ &
$v^2$ & $\kappa_\textrm{eff}(v)$ & $-S_K^{\prime} S_V \xi$ &  \\ %
\hline
$<0$ & \ $-$ \ & $(1,2)$&      $<-2$ & $-$ & $+\kappa$ & $-\kappa$ & $-$ & \textsf{X} \\
        &             & &  $(-2,-y)$ & $+$ & $+\kappa$ & $+\kappa$ & $+$ & \textsf{O} \\
        &             & &  $(-y,-1)$ & $-$ & $+\kappa$ & $+\kappa$ & $-$ & \textsf{X} \\
        &             & &      $>-1$ & $-$ & $-\kappa$ & $+\kappa$ & $-$ & \textsf{X} \\
\hline
$(0,\frac{1}{6})$ & \ $+$ \ &$>2$&$<1$&$-$ & $+\kappa$ & $+\kappa$ & $+$ & \textsf{O} \\
        &             & &   $(1,2)$  & $-$ & $-\kappa$ & $+\kappa$ & $+$ & \textsf{X} \\
        &             & &   $(2,y)$  & $+$ & $-\kappa$ & $-\kappa$ & $-$ & \textsf{X} \\
        &             & &      $>y$  & $-$ & $-\kappa$ & $-\kappa$ & $+$ & \textsf{O} \\
\hline
$>\frac{1}{6}$ & \ $+$ \ &$<1$& $<y$ & $+$ & $+\kappa$ & $+\kappa$ & $-$ & \textsf{X} \\
        &             & &    $(y,1)$ & $-$ & $+\kappa$ & $+\kappa$ & $+$ & \textsf{O} \\
        &             & &    $(1,2)$ & $-$ & $-\kappa$ & $+\kappa$ & $+$ & \textsf{X} \\
        &             & &       $>2$ & $+$ & $-\kappa$ & $-\kappa$ & $-$ & \textsf{X} \\
\multicolumn{9}{l}{$S_V = -1$} \\
$\xi$ & $A_M$ & $y$ & $A_0/|A_M|$ & $S_K^{\prime}$ &
$v^2$ & $\kappa_\textrm{eff}(v)$ & $-S_K^{\prime} S_V \xi$ &  \\ %
\hline
$<0$ & \ $+$ \ & $(1,2)$&      $<1$  & $-$ & $-\kappa$ & $+\kappa$ & $+$ & \textsf{X} \\
        &             & &   $(1,y)$  & $-$ & $+\kappa$ & $+\kappa$ & $+$ & \textsf{O} \\
        &             & &   $(y,2)$  & $+$ & $+\kappa$ & $+\kappa$ & $-$ & \textsf{X} \\
        &             & &      $>2$  & $-$ & $+\kappa$ & $-\kappa$ & $+$ & \textsf{X} \\
\hline
$(0,\frac{1}{6})$ & \ $-$\ &$>2$&$<-y$&$-$ & $-\kappa$ & $-\kappa$ & $-$ & \textsf{X} \\
        &             & &  $(-y,-2)$ & $+$ & $-\kappa$ & $-\kappa$ & $+$ & \textsf{O} \\
        &             & &  $(-2,-1)$ & $-$ & $-\kappa$ & $+\kappa$ & $-$ & \textsf{X} \\
        &             & &      $>-1$ & $-$ & $+\kappa$ & $+\kappa$ & $-$ & \textsf{X} \\
\hline
$>\frac{1}{6}$ & \ $-$ \ &$<1$&$<-2$ & $+$ & $-\kappa$ & $-\kappa$ & $+$ & \textsf{O} \\
        &             & &  $(-2,-1)$ & $-$ & $-\kappa$ & $+\kappa$ & $-$ & \textsf{X} \\
        &             & &  $(-1,-y)$ & $-$ & $+\kappa$ & $+\kappa$ & $-$ & \textsf{X} \\
        &             & &      $>-y$ & $+$ & $+\kappa$ & $+\kappa$ & $+$ & \textsf{O} \\ %
\end{tabular}
\end{table*}



\begin{table*}
\caption{\label{table:feasible cases} Feasible cases. The name of
each feasible case is given in the first column ``ID'', where
``P'' and ``Q'' mean phantom and quintessence respectively and the
information about the sign or the range of the essential
quantities (for specifying a model), $\kappa$, $V$ and $\xi$, are
presented in the subscript in order. The sign of the resulting
effective CC and the extent of the automatic CC tuning in these
feasible cases are shown in the columns $A_M$ and $A_0/|A_M|$. The
cases where the original bare CC under control can be arbitrarily
large in its magnitude are particularly interesting and marked by
the star symbol ``$\star$''.}
\begin{tabular}{r|ccc|cc}
\multicolumn{6}{l}{\hspace{-0.5em} Quintessence ($S_K = +1$)} \vspace{0.2em}    \\
       ID \ \ & $\kappa$ &  $V$  & $\xi$ &    $A_M$ & $A_0/|A_M|$ \\
\hline
        Q$_{++-}$ &      $+$ &  $+$  &   $-$ &      $-$ &  $(-2,-1)$ \\
$^\star$Q$_{+-+}$ &      $+$ &  $-$  &   $+$ &      $-$ &   $ > -1 $ \\
$^\star$Q$_{--+}$ &      $-$ &  $-$  &   $+$ &      $-$ &   $ < -2 $ \\ %
\multicolumn{6}{c}{} \\
\multicolumn{6}{c}{}
\end{tabular}
\hspace{1em}
\begin{tabular}{r|ccc|cc|r}
\multicolumn{7}{l}{\hspace{-0.5em} Phantom ($S_K = -1$)}  \vspace{0.2em}  \\
           ID \ \ & $\kappa$ & $V$ &             $\xi$ & $A_M$ & $A_0/|A_M|$ & {\footnotesize (remark)} \\
\hline
        P$_{++1}$ &      $+$ & $+$ &              $<0$ &      $-$ &    $(-2,-y)$ & $1<y<2$ \\
$^\star$P$_{++2}$ &      $+$ & $+$ & $(0,\frac{1}{6}]$ &      $+$ &      $ < 1 $ &         \\
        P$_{++3}$ &      $+$ & $+$ &    $>\frac{1}{6}$ &      $+$ &      $(y,1)$ &   $y<1$ \\ %
        P$_{+-1}$ &      $+$ & $-$ &              $<0$ &      $+$ &      $(1,y)$ & $1<y<2$ \\
$^\star$P$_{+-3}$ &      $+$ & $-$ &    $>\frac{1}{6}$ &      $-$ &     $ > -y $ &   $y<1$ %
\end{tabular}
\hspace{0.2em}
\begin{tabular}{r|ccc|cc|c}
\multicolumn{7}{l}{}  \vspace{0.2em} \\
           ID \ \ & $\kappa$ & $V$ &             $\xi$ & $A_M$ & $A_0/|A_M|$ & {\footnotesize (remark)} \\
\hline
$^\star$P$_{-+2}$ &      $-$ & $+$ & $(0,\frac{1}{6})$ &      $+$ &      $ > y $ &   $y>2$ \\
        P$_{--2}$ &      $-$ & $-$ & $(0,\frac{1}{6})$ &      $-$ &    $(-y,-2)$ &   $y>2$ \\
$^\star$P$_{--3}$ &      $-$ & $-$ & $\ge \frac{1}{6}$ &      $-$ &      $ <-2 $ &   $y<1$ \\ %
\multicolumn{6}{c}{} \\
\multicolumn{6}{c}{}
\end{tabular}
\end{table*}


In the limit $|A_0/A_M| \gg 1$, i.e., when the CC tuning involves
significant CC reduction,
\begin{eqnarray}
v^2 &\simeq& - (A_0/A_M) \cdot (\kappa \xi)^{-1} \, , \\
\kappa_\textrm{eff}(v) &\simeq& - (A_0/A_M)^{-1} \cdot \kappa
\simeq (\xi v^2)^{-1} \, .
\end{eqnarray}
Requiring $\kappa_\textrm{eff} \sim M_\textrm{Pl}^{-2}$, one
obtains
\begin{eqnarray}
v^2 &\sim& M_\textrm{Pl}^2 / \xi \, , \label{eq:v vs Mpl} \\
\kappa &\sim& - (A_0/A_M) \cdot M_\textrm{Pl}^{-2} \, .
\label{eq:kappa vs Mpl}
\end{eqnarray}
According to Eq.\ (\ref{eq:v vs Mpl}), for this automatic CC
control mechanism to work for arbitrarily large bare CC, the
non-minimal coupling constant $\xi$ must be positive, which is
exhibited in the $\star$ cases. Eq.\ (\ref{eq:kappa vs Mpl})
indicates an important feature that through the process of the CC
reduction under this mechanism, gravity is weakened by a factor
$A_M/A_0$.

The change of the gravitational interaction strength involved in
the CC tuning under this mechanism is a general feature. Regarding
the strength of gravity and the scale of the CC, several
interesting cases are demonstrated in Table \ref{table:scales},
and the discussions about the examples therein are as follows.
\vspace{0.5em} \\
\textsl{Case} (1): All are around the Planck scale, benefiting the
RS scenario where it is required to control the effective CCs on
the branes and that in the bulk to some fixed values
around the Planck scale. \vspace{0.5em} \\ %
\textsl{Case} (2): The original strength of gravity is on the TeV
scale, and accordingly the hierarchy between gravity and three
gauge interactions in the standard model of particle physics is
bridged at the very beginning. The third example in this category
is particularly interesting, where both gravity and the CC are on
the TeV scale originally. This TeV-scale gravity eventually
becomes the present Planck-scale one through the process of
reducing an originally TeV-scale CC to the scale of $10^{-4}$eV.
Note that the resulting effective CC is barely affected by the
later CC change(s) as long as the scale the CC change is much
smaller than TeV. That is, this CC control is stable against the
CC change up to the TeV scale.
\vspace{0.5em} \\ %
\textsl{Case} (3): For tuning a large CC to the $H_0$-scale value
suggested by observations, the gravity must be very strong
originally, which leads to a fine tuning of the original strength
of gravity.


\begin{table}
\caption{\label{table:scales} Interesting cases regarding the
energy scales of $A_0$ ($\Lambda_0$), $A_M$
($\Lambda_\textrm{eff}$) and $\kappa$ (the original strength of
gravity).}
\begin{ruledtabular}
\begin{tabular}{lcr}
$A_0^{1/2}$ & $A_M^{1/2}$ & $\kappa^{-1/2}$ \\
\hline %
            & $A_0^{1/2}$ & $M_\textrm{Pl}$ \\
            & $10^{-3} A_0^{1/2}$ & $M_\textrm{GUT}$\footnotemark[1] \\%
            & $10^{-16} A_0^{1/2}$ & TeV\footnotemark[2] \\
            & $A_0 / M_\textrm{Pl}$ & $A_0^{1/2}$\footnotemark[3] \\
\hline %
\multicolumn{3}{l}{(1) (all Planck-scale)} \\
$M_\textrm{Pl}$ & $M_\textrm{Pl}$\footnotemark[4] & $M_\textrm{Pl}$ \\
\hline
\multicolumn{3}{l}{(2) (TeV gravity)\footnotemark[2]} \\
$M_\textrm{Pl}$ & TeV & TeV\footnotemark[2] \\
$M_\textrm{GUT}$ & GeV & TeV\footnotemark[2] \\
TeV & $10^{-4}$eV & TeV\footnotemark[2]\footnotemark[3] \\
\hline
\multicolumn{3}{l}{(3) ($\Lambda_\textrm{eff} \sim H_0^2$)\footnotemark[5]} \\
TeV & $H_0$\footnotemark[5] & $10^{-17}$eV\footnotemark[6] \\
$M_\textrm{Pl}$ & $H_0$\footnotemark[5] & $H_0$\footnotemark[6] \\
\end{tabular}
\end{ruledtabular}
\footnotetext[1]{$M_\textrm{GUT} \sim 10^{16}$GeV.} %
\footnotetext[2]{TeV-scale gravity originally.} %
\footnotetext[3]{The same as the scale of the original bare CC.} %
\footnotetext[4]{Planck-scale effective CC as required in the RS scenario.}%
\footnotetext[5]{The current CC scale suggested by observations, $H_0 \sim 10^{-33}$eV.} %
\footnotetext[6]{Extremely strong gravity originally.} %
\end{table}



\textit{Summary.} It has been a task for a long time to explain
the smallness of the CC against the huge contribution from the
quantum fluctuations of vacuum and the possible significant change
during the SSB phase transition. In the present article a
mechanism to automatically control the CC via non-minimally
coupled phantom and quintessence is proposed. While the
quintessence can only direct the CC to a negative value, the
phantom can generate an effective CC of some required size, either
positive or negative, from an arbitrarily large bare CC. With this
mechanism, for the RS scenario controlling the CCs on the branes
and that in the bulk to some fixed values around the Planck scale
can be achieved. In addition, controlling the CC to a value on the
$10^{-4}$eV scale or above, such as TeV, is also achievable.
In the case involving significant CC reduction, gravity is
weakened and accordingly should be stronger originally. There
exist the cases where the gravity is on the TeV scale originally.
This is particularly interesting because in these cases gravity
and three gauge interactions in the standard model of particle
physics can be on the equal footing, i.e.\ with no hierarchy
between them, at the very beginning. Furthermore, in one the these
cases, the original bare CC is also on the TeV scale that
accordingly plays a particularly fundamental role in the
beginning. Later the four fundamental interactions split into
gravity and other three gauge interactions with the hierarchy of
the 16 orders of magnitude, as accompanied by the reduction of the
CC from the TeV scale to the $10^{-4}$eV scale.
Although the CC can be automatically controlled under this
mechanism to a large extent, generating an effective CC on the
observation-suggested $H_0$ scale is still a difficult mission yet
to be completed.

\textsf{Acknowledgments}. This work is supported by the Taiwan
National Science Council (NSC 96-2119-M-007-001).



\begin{thebibliography}{00}

\bibitem{Perlmutter:1998np}
  S.~Perlmutter {\it et al.}  [Supernova Cosmology Project Collaboration],
  Astrophys.\ J.\  {\bf 517}, 565 (1999).

\bibitem{Riess:1998cb}
  A.~G.~Riess {\it et al.}  [Supernova Search Team Collaboration],
  Astron.\ J.\  {\bf 116}, 1009 (1998).

\bibitem{SNLS}
  P.~Astier {\it et al.}  [SNLS Collaboration],
  {\it A\&A} {\bf 447}, 31 (2006).

\bibitem{Riess:2006fw}
  A.~G.~Riess {\it et al.},
  Astrophys.\ J.\  {\bf 659}, 98 (2007).

\bibitem{ESSENCE:2007jb}
  W.~M.~Wood-Vasey {\it et al.}  [ESSENCE Collaboration],
  Astrophys.\ J.\  {\bf 666}, 694 (2007).

\bibitem{WMAP:2006}
  D.~N.~Spergel {\it et al.}  [WMAP Collaboration],
  Astrophys.\ J.\ Suppl.\  {\bf 170}, 377 (2007).

\bibitem{Tegmark:2006}
  M.~Tegmark {\it et al.},
  Phys.\ Rev.\  D {\bf 74}, 123507 (2006).

\bibitem{Weinberg:1988cp}
  S.~Weinberg,
  Rev.\ Mod.\ Phys.\  {\bf 61}, 1 (1989).

\bibitem{Padmanabhan:2002ji}
  T.~Padmanabhan,
  Phys.\ Rept.\  {\bf 380}, 235 (2003).

\bibitem{Nobbenhuis:2004wn}
  S.~Nobbenhuis,
  Found.\ Phys.\  {\bf 36}, 613 (2006).

\bibitem{RS:1999}
  L.~Randall and R.~Sundrum,
  Phys.\ Rev.\ Lett.\  {\bf 83}, 3370 (1999);
%
  {\bf 83}, 4690 (1999).

\bibitem{Gu:2001rr}
  Je-An~Gu and W-Y.~P.~Hwang,
  Mod.\ Phys.\ Lett.\  A {\bf 17}, 1979 (2002).

\end{thebibliography}
\end{document}